\documentclass[letterpaper, 10 pt, conference]{ieeeconf}
\usepackage{amsfonts}
\usepackage{amsmath}
\usepackage{amssymb}
\usepackage{color}
\usepackage{cite}
\usepackage{epsfig}
\usepackage{graphics}
\usepackage{graphicx}
\usepackage{ifpdf}
\usepackage{subfigure}
\usepackage{times}

\IEEEoverridecommandlockouts                              
\newtheorem{theorem}{Theorem}

\newtheorem{corollary}[theorem]{Corollary}

\newtheorem{example}[theorem]{Example}

\newtheorem{lemma}[theorem]{Lemma}

\newtheorem{proposition}[theorem]{Proposition}
\newtheorem{remark}[theorem]{Remark}

\begin{document}

\title{{\LARGE \textbf{Analysis of Equilibria and Strategic Interaction in
Complex Networks}}}
\author{Victor M. Preciado, Jaelynn Oh and Ali Jadbabaie \thanks{%
This work is supported by ONR MURI N000140810747.} \thanks{%
V.M. Preciado and A. Jadbabaie are with the GRASP Laboratory, School of
Engineering and Applied Science, University of Pennsylvania, Philadelphia,
PA 19104, USA \texttt{\footnotesize \{preciado,jadbabai\}@seas.upenn.edu}.
J. Oh is with the Department of Operations and Information Management,
Wharton School, University of Pennsylvania, Philadelphia, PA 19104, USA 
\texttt{\small jaelynn@wharton.upenn.edu}}}
\maketitle

\begin{abstract}
This paper studies $n$-person simultaneous-move games with linear best
response function, where individuals interact within a given network
structure. This class of games have been used to model various settings,
such as, public goods, belief formation, peer effects, and oligopoly. The
purpose of this paper is to study the effect of the network structure on
Nash equilibrium outcomes of this class of games. Bramoull\'{e} et al.
derived conditions for uniqueness and stability of a Nash equilibrium in
terms of the smallest eigenvalue of the adjacency matrix representing the
network of interactions. Motivated by this result, we study how local
structural properties of the network of interactions affect this eigenvalue,
influencing game equilibria. In particular, we use algebraic graph theory
and convex optimization to derive new bounds on the smallest eigenvalue in
terms of the distribution of degrees, cycles, and other relevant
substructures. We illustrate our results with numerical simulations
involving online social networks.
\end{abstract}

\section{Introduction}

In most social and economic settings, individuals do not interact uniformly
with the rest of a society. Instead, they influence each other according to
a structured network of interactions. This network can represent friendships
in a social network, transactions among firms in a market, or communication
links in a process of belief formation. In this context, an interesting
question is to study how the network structure affects the outcome of the
interactions of the agents. With this purpose, one can model strategic
interactions in a networked society as a multi-player simultaneous-move
game. In particular, we focus our attention on the broad class of games with
linear best response function \cite{FT91}. This class of games have been
used to model various settings such as belief formation \cite{AP07}, peer
effects \cite{BCZ06}, and public goods \cite{BK07}.

In order to analyze the influence of the network structure on the game
outcome, we use two recent results by Bramoull\'{e} et al. \cite{BKA10} and
Ballester et al. \cite{BC07} relating the Nash equilibria of the game with
the largest and smallest eigenvalues of the adjacency matrix of
interactions. For example, in \cite{BKA10}, the authors show that uniqueness
and stability of a Nash equilibrium in games with linear best responses can
be determined by the smallest eigenvalue of the network. In \cite{BKA10}, it
was also illustrated how the smallest eigenvalue of the adjacency matrix
determines the capacity of the network to absorb perturbations on the
actions of the agents.

Motivated by the results of Bramoull\'{e} et al. \cite{BKA10}, we study how
local structural properties of the network affect the smallest eigenvalue of
the adjacency matrix of interactions, affecting game equilibria. Therefore,
our results build a bridge between structural properties of a network of
interactions and the outcome of games with linear best responses. In
particular, we use algebraic graph theory and convex optimization to derive
bounds on the smallest eigenvalue of the adjacency matrix in terms of the
distribution of degrees, cycles, and other important substructures. As we
illustrate with numerical simulations in online social networks, these
bounds can be used to estimate the effect of structural perturbations on the
smallest eigenvalue. 

The paper is organized as follows. In the next subsection, we review
graph-theoretical terminology needed in our derivations. In Section \ref%
{Strategic Section}, we review the relationship between the equilibria of
games with linear best responses in a network and the smallest eigenvalue of
the adjacency matrix of interactions. In Section~\ref{Spectral Section}, we
use algebraic graph theory to derive closed-form expressions for the
so-called spectral moments of a network in terms of local structural
features. In Section \ref{Optimal Bounds}, we use convex optimization to
derive optimal bounds on the smallest (and largest) eigenvalue of the
interaction network from these moments. Our bounds help us to understand how
structural properties of a network impact the stability properties of the
Nash equilibria in the game. We illustrate our results with numerical
simulations in real online social networks in Section \ref{Simulations}.

\subsection{Notation}

Let $\mathcal{G}=\left( \mathcal{V},\mathcal{E}\right) $ denote an
undirected graph with $n$ nodes, $e$ edges, and no self-loops\footnote{%
An undirected graph with no self-loops is also called a \emph{simple} graph.}%
. We denote by $\mathcal{V}\left( \mathcal{G}\right) =\left\{ v_{1},\dots
,v_{n}\right\} $ the set of nodes and by $\mathcal{E}\left( \mathcal{G}%
\right) \subseteq \mathcal{V}\left( \mathcal{G}\right) \times \mathcal{V}%
\left( \mathcal{G}\right) $ the set of undirected edges of $\mathcal{G}$. If 
$\left\{ v_{i},v_{j}\right\} \in \mathcal{E}\left( \mathcal{G}\right) $ we
call nodes $v_{i}$ and $v_{j}$ \emph{adjacent} (or neighbors), which we
denote by $v_{i}\sim v_{j}$ and define the set of neighbors of $v_{i}$ as $%
\mathcal{N}_{i}=\{w\in \mathcal{V}\left( \mathcal{G}\right) :\left\{
v_{i},w\right\} \in \mathcal{E}\left( \mathcal{G}\right) \}$. The number of
neighbors of $v_{i}$ is called the \emph{degree} of the node, denoted by $%
d_{i}$. We define a \emph{walk} of length $k$ from $v_{0}$ to $v_{k}$ to be
an ordered sequence of nodes $\left( v_{0},v_{1},...,v_{k}\right) $ such
that $v_{i}\sim v_{i+1}$ for $i=0,1,...,k-1$. If $v_{0}=v_{k}$, then the
walk is closed. A closed walk with no repeated nodes (with the exception of
the first and last nodes) is called a \emph{cycle}. For example, \emph{%
triangles}, \emph{quadrangles} and \emph{pentagons} are cycles of length
three, four, and five, respectively.

Graphs can be algebraically represented via matrices. The adjacency matrix
of an undirected graph $\mathcal{G}$, denoted by $A_{\mathcal{G}}=[a_{ij}]$,
is an $n\times n$ symmetric matrix defined entry-wise as $a_{ij}=1$ if nodes 
$v_{i}$ and $v_{j}$ are adjacent, and $a_{ij}=0$ otherwise\footnote{%
For simple graphs, $a_{ii}=0$ for all $i$.}. The eigenvalues of $A_{\mathcal{%
G}}$, denoted by $\lambda _{1}\geq \lambda _{2}\geq \ldots \geq \lambda _{n}$%
, play a key role in our paper. The spectral radius of $A_{\mathcal{G}}$,
denoted by $\rho \left( A_{\mathcal{G}}\right) $, is the maximum among the
magnitudes of its eigenvalues. Since $A_{\mathcal{G}}$ is a symmetric matrix
with nonnegative entries, all its eigenvalues are real and the spectral
radius is equal to the largest eigenvalue, $\lambda _{1}$. We define the $k$%
-th spectral moment of the adjacency matrix $A_{\mathcal{G}}$ as:%
\begin{equation}
m_{k}\left( A_{\mathcal{G}}\right) =\frac{1}{n}\sum_{i=1}^{n}\lambda
_{i}^{k}.  \label{Spectral Moment}
\end{equation}%
As we shall show in Section \ref{Spectral Section}, there is a direct
connection between the spectral moments and the presence of certain
substructures in the graph, such as cycles of length $k$.

%%%%%%%%%%%%%%%%%%%%%%%%%%%%%%%%%%%%%%%%%%%%%%%%%%%%%%%%%%%%%%%%%%%%%%%%%

\section{Strategic Interactions in Networks\label{Strategic Section}}

In this section we present the game-theoretical model of strategic
interactions considered in this paper and present interesting connections
between the Nash equilibria and the eigenvalues of the adjacency matrix of
the network.

\subsection{The Model}

We represent the network of influences using a simple graph $\mathcal{G}$.
Let $\mathcal{N}=\{1,\cdots ,n\}$ denote the set of $n$ players located at
each node of the graph $\mathcal{G}$. We denote by $x_{i}\in \lbrack
0,\infty )$ the action chosen by agent $i$, and by $\mathbf{x}$ the vector
that represents the joint actions for all agents. We denote by $\mathbf{x}%
_{-i}$ the vector of actions of all players excluding player $i$. As
mentioned before, players interact according to a network of influences that
we describe using its adjacency matrix $A_{\mathcal{G}}$. The interactions
are assumed to be symmetric, $a_{ij}=a_{ji}$, and we do not allow
self-loops, $a_{ii}=0$. The payoff function for agent $i$ is given by:%
\begin{equation*}
U_{i}(x_{i},\mathbf{x}_{-i};\delta ,A_{\mathcal{G}})
\end{equation*}%
where $\delta \in \mathbb{R}$ is a parameter that can be tuned to change the
influence of neighboring nodes on each player's action.

\subsection{\label{Game Description}Games with Linear Best Response Functions%
}

We study a class of games whose best response functions take a linear form.
One well known example of this class of games is the differentiated-product
Cournot oligopoly with linear inverse demand and constant marginal cost with
payoff function defined as \cite{BKA10}:%
\begin{equation}
U_{i}(x_{i},\mathbf{x}_{-i};\delta ,A_{\mathcal{G}})=x_{i}\left(
a-b(x_{i}+2\delta \sum_{j=1}^{n}a_{ij}x_{j})\right) -dx_{i},
\label{Game Payoffs}
\end{equation}%
where $d$ is the constant marginal cost, and $x_{i}$ represents the amount
produced by agent $i$ in the oligopoly. Here, the inverse demand for agent $i
$ is given by $P_{i}(x_{i},\mathbf{x}_{-i};\delta ,A_{\mathcal{G}%
})=a-b(x_{i}+2\delta \sum_{j=1}^{n}a_{ij}x_{j})$. One can prove that the
best response function for this type of games yield the form \cite{BKA10}:%
\begin{equation}
f_{i}(\mathbf{x},\delta ,A_{\mathcal{G}})=\max (0,\bar{x}_{i}-\delta
\sum_{j=1}^{n}a_{ij}x_{j}),  \label{Best Reply Function}
\end{equation}%
where $\bar{x}_{i}$ is the action that agent $i$ would take in isolation,
i.e., $\bar{x}_{i}\in \arg \max_{x_{i}}U_{i}(x_{i},\mathbf{x}_{-i};\delta
,A_{\mathcal{G}})$ with $a_{ij}=0$, \ for all $j$. Without loss of
generality, one can normalize $\bar{x}_{i}\equiv 1$ for all $i$, so that $%
f_{i}(\mathbf{x})\in \lbrack 0,1]$. Then, a Nash equilibrium for this game
is a vector $\mathbf{x}\in \lbrack 0,1]^{n}$ that satisfies $x_{i}=f_{i}(%
\mathbf{x},\delta ,A_{\mathcal{G}})$, for all agents $i\in \mathcal{N}$,
simultaneously. In what follows, we briefly describe a strategy to compute
the complete set of Nash equilibria for $\delta \in \lbrack 0,1]$.

\subsection{Complete Set of Nash Equilibria}

Using the best response function in (\ref{Best Reply Function}), one can
determine the entire set of equilibria by simultaneously solving for the
best response of each player. In \cite{BKA10}, an algorithm that finds the
full set of Nash equilibria in exponential time is proposed. For a vector $%
\mathbf{x}$, let $S$ denote the set of active agents, i.e., $S=\{i:x_{i}>0\}$%
. Let $\mathbf{x}_{S}$ denote the vector of actions of the agents in $S$.
The set of active players induce a subgraph $\mathcal{G}_{S}\subseteq 
\mathcal{G}$, with node-set $S\subseteq \mathcal{V}\left( \mathcal{G}\right) 
$ and a set of edges $\mathcal{E}_{S}\subseteq \mathcal{E}\left( \mathcal{G}%
\right) $ connecting active agents. We denote by $\mathcal{G}_{\mathcal{N}%
\backslash S,S}$ the subgraph of $\mathcal{G}$ whose edges connect active
agents in $S$ to inactive agents in $\mathcal{N}\backslash S$. The adjacency
matrices of $\mathcal{G}_{S}$ and $\mathcal{G}_{\mathcal{N}\backslash S,S}$
are denoted by $A_{S}\in \mathbb{R}^{\left\vert S\right\vert \times
\left\vert S\right\vert }$ and $A_{\mathcal{N}\backslash S,S}\in \mathbb{R}%
^{n-\left\vert S\right\vert \times \left\vert S\right\vert }$, respectively.
Then, one can show the following \cite{BKA10}:

\begin{proposition}
\label{Nash Conditions}A profile $\mathbf{x}$ with active agents $S$ is a
Nash equilibrium if and only if: 
\begin{eqnarray*}
(I_{\left\vert S\right\vert }+\delta A_{S})\mathbf{x}_{S} &=&\mathbf{1}%
_{\left\vert S\right\vert }\mathbf{,}\text{ and} \\
\delta A_{\mathcal{N}\backslash S,S}~\mathbf{x}_{S} &\geq &\mathbf{1}%
_{n-\left\vert S\right\vert }\mathbf{,}
\end{eqnarray*}%
where $I_{p}$ is the $p\times p$ identity matrix and $\mathbf{1}_{q}$ is the 
$q$-dimensional vector of ones.
\end{proposition}

\bigskip

Thus, in order to determine the complete set of all Nash equilibria, one can
check the conditions in Proposition \ref{Nash Conditions} for each one of
the $2^{n}$ possibilities of $S$. For each possible $S$, these conditions
can be checked by computing $\mathbf{x}_{S}=(I_{\left\vert S\right\vert
}+\delta A_{S})^{-1}\mathbf{1}_{\left\vert S\right\vert }$,\footnote[1]{%
The matrix $\left( I_{\left\vert S\right\vert }+\delta A_{S}\right) $ is
invertible for almost any $\delta \in R$, excepting the measure zero set $%
\{-1/\mu _{i}$, $i=1,...,\left\vert S\right\vert \}$, where $\mu
_{1},...,\mu _{\left\vert S\right\vert }$ are the eigenvalues of $A_{S}$.}
and checking whether $\delta A_{\mathcal{N}\backslash S,S}~\mathbf{x}%
_{S}\geq \mathbf{1}_{n-\left\vert S\right\vert }$. If the last inequality
holds, then $\mathbf{x}_{S}$ is an equilibrium outcome. Note that using this
approach to compute the set of equilibria runs in exponential time. However,
we can analyze some properties of the Nash equilibria, such as uniqueness
and stability, by looking into the eigenvalues of the adjacency matrix.

\subsection{The Shape of Nash Equilibria}

In order to relate the equilibrium outcomes of the game to the network
structure, the authors in \cite{BKA10} defined the following potential
function:%
\begin{equation*}
\varphi (\mathbf{x};\delta ,A_{\mathcal{G}})\triangleq \sum_{i=1}^{n}\left(
x_{i}-\frac{1}{2}x_{i}^{2}\right) -\frac{1}{2}\delta
\sum_{i=1}^{n}\sum_{j=1}^{n}a_{ij}x_{i}x_{j}.
\end{equation*}%
Then, they proved, using Kuhn-Tucker conditions, that the set of Nash
equilibria coincides with the critical points of the following optimization
problem:%
\begin{equation*}
\begin{array}{lll}
(\text{P}) & \max_{\mathbf{x}} & \varphi (\mathbf{x};\delta ,A_{\mathcal{G}})
\\ 
& s.t. & x_{i}\geq 0,\text{ for all }i,%
\end{array}%
\end{equation*}%
for a given network structure $A_{\mathcal{G}}$ and a parameter $\delta $.

\subsection{Eigenvalues and Nash Equilibria}

We can find several results in the literature providing sufficient
conditions for the existence of a unique Nash equilibrium in games with
linear best response functions in terms of the eigenvalues of the network of
influences. We enumerate below some sufficient conditions that are related
with our work:

\begin{proposition}
\label{Kranton condition}Consider the class of games with linear best
response functions described in Section \ref{Game Description}. For these
games, we have the following sufficient conditions for the existence of a
unique Nash equilibrium:

\begin{description}
\item[(\emph{i})] $\delta <-1/\lambda _{n}(A_{\mathcal{G}})$, (Bramoull\'{e}
et al., \cite{BKA10}).

\item[(\emph{ii})] $\delta <1/\rho \left( A_{\mathcal{G}}\right) $,
(Ballester et al., \cite{BC07}).
\end{description}
\end{proposition}

\bigskip

We can compare the set of spectral conditions in Proposition \ref{Kranton
condition} using the following inequalities \cite{BKA10}:

\begin{lemma}
For any simple graph $\mathcal{G}$, we have that $-1/\lambda _{n}(A_{%
\mathcal{G}})\geq 1/\rho \left( A_{\mathcal{G}}\right) $, where this
inequality is strict when no component of $\mathcal{G}$ is bipartite.
\end{lemma}

\begin{remark}
\label{Active Remark}Hence, Condition (\emph{i}) in Proposition \ref{Kranton
condition} provides the best sufficient condition for the uniqueness of Nash
equilibrium in these games. Furthermore, one can also prove that under
Condition (i) or (ii), all players are active at the equilibrium point,
i.e., $S=\mathcal{N}$ \cite{BKA10}.
\end{remark}

In Section \ref{Optimal Bounds}, we shall derive upper bounds on $\lambda
_{n}\left( A_{\mathcal{G}}\right) $ in terms of structural properties of the
network. These bounds, in combination with Condition (\emph{i}) in
Proposition \ref{Kranton condition}, will allow us to derive sufficient
conditions for the existence of a Nash equilibrium in terms of structural
properties of the network.

\bigskip

\subsection{The Stability of Nash Equilibria}

We present conditions for stability of a Nash equilibrium in terms of $%
\lambda _{n}(A_{S})$. A Nash equilibrium $\mathbf{x}$ is asymptotically
stable when the system of differential equations:%
\begin{equation*}
\begin{array}{c}
\dot{x}_{1}=h_{1}(\mathbf{x})=f_{1}(\mathbf{x};\delta ,A_{\mathcal{G}%
})-x_{1}, \\ 
\vdots  \\ 
\dot{x}_{n}=h_{n}(\mathbf{x})=f_{n}(\mathbf{x};\delta ,A_{\mathcal{G}%
})-x_{n},%
\end{array}%
\end{equation*}%
is locally asymptotically stable around $\mathbf{x}$. One can prove the
following necessary and sufficient condition for an equilibrium $\mathbf{x}$
to be asymptotically stable \cite{BKA10}:

\begin{lemma}
An equilibrium profile $\mathbf{x}$ is asymptotically stable if and only if $%
\delta <-1/\lambda _{n}(A_{S})$ and $\delta \sum_{j=1}^{n}a_{ij}x_{j}>1,$
for all inactive agents $i\in \mathcal{N}\backslash S$.
\end{lemma}

\bigskip

From the above lemma and Remark \ref{Active Remark}, we conclude that if $%
\delta <-1/\lambda _{n}(A_{\mathcal{G}})$, there is a unique Nash
equilibrium and it is asymptotically stable. These results show the close
connection between the smallest and the largest eigenvalues of the adjacency
matrix of interactions and the outcome of games with linear best response
functions in a network. On the other hand, the results in this section are
applicable if we are able to compute the eigenvalues of $A_{\mathcal{G}}$.
For many large-scale complex networks, the structure of the network can be
very intricate \cite{S01}-\cite{Pre08} ---in many cases not even known
exactly--- and an explicit eigenvalue decomposition can be very challenging
to compute, if not impossible. In many practical settings, instead of having
access to the complete network topology, we have access to local
neighborhoods of the network structure around a set of nodes. In this
context, it is important to understand the impact of local structural
information on the eigenvalues of the adjacency matrix. In the rest of the
paper, we propose a novel methodology to compute optimal bounds on relevant
eigenvalues of $A_{\mathcal{G}}$ from local information regarding the
network structure. Our results allow us to study the role of local
structural information in the outcome of games with linear best response
functions.

\bigskip

\section{Spectral Analysis of the Interaction Matrix\label{Spectral Section}}

We study the relationship between a network's local structural properties
and the smallest eigenvalue of its adjacency matrix. Algebraic graph theory
provides us with tools to relate the eigenvalues of a network with its
structural properties. Particularly useful is the following well-known
result relating the $k$-th spectral moment of $A_{\mathcal{G}}$ with the
number of closed walks of length $k$ in $\mathcal{G}$ \cite{Big93}:

\begin{lemma}
\label{Moments from Walks}Let $\mathcal{G}$ be a simple graph. The $k$-th
spectral moment of the adjacency matrix of $\mathcal{G}$ can be written as%
\begin{equation}
m_{k}(A_{\mathcal{G}})=\frac{1}{n}\sum_{i=1}^{n}\lambda_{i}^{k}=\frac{1}{n}%
\left\vert \Psi_{\mathcal{G}}^{\left( k\right) }\right\vert ,
\label{Moments as Walks in Graph}
\end{equation}
where $\Psi_{\mathcal{G}}^{\left( k\right) }$ is the set of all closed walks
of length $k$ in $\mathcal{G}$. \footnote{%
We denote by $\left\vert Z\right\vert $ the cardinality of a set $Z$.}
\end{lemma}

\medskip

From (\ref{Moments as Walks in Graph}), we can easily compute the first
three spectral moments of $A_{\mathcal{G}}$ in terms of the number of nodes,
edges and triangles as follows \cite{Big93}:

\begin{corollary}
\label{One corollary}Let $\mathcal{G}$ be a simple graph with adjacency
matrix $A_{\mathcal{G}}$. Denote by $n$, $e$ and $\Delta $ the number of
nodes, edges and triangles in $\mathcal{G}$, respectively. Then,%
\begin{align}
m_{1}(A_{\mathcal{G}})& =0,  \label{Moments as Averages} \\
m_{2}(A_{\mathcal{G}})& =2e/n,  \notag \\
m_{3}(A_{\mathcal{G}})& =6\,\Delta /n.  \notag
\end{align}
\end{corollary}

\bigskip

\begin{remark}
Notice that the coefficients 2 (resp. 6) in the above expressions
corresponds to the number of closed walks of length 2 (resp. 3) enabled by
the presence of an edge (resp. triangle). Similar expressions can be derived
for higher-order spectral moments, although a more elaborated combinatorial
analysis in required.
\end{remark}

In our case, we are interested in the following expressions, derived in \cite%
{PJ10}, for the first five spectral moments of $\mathcal{G}$:

\begin{lemma}
\label{Lemma 4th metrics}Let $\mathcal{G}$ be a simple graph. Denote by $e$, 
$\Delta $, $Q$ and $\Pi $ the total number of edges, triangles, quadrangles
and pentagons in $\mathcal{G}$, respectively. Define $W_{2}=%
\sum_{i=1}^{n}d_{i}^{2}$ and $\mathcal{C}_{dt}=\sum_{i}d_{i}t_{i}$, where $%
t_{i}$ is the number of triangles touching node $i$. Then,%
\begin{eqnarray}
m_{4}\left( A_{\mathcal{G}}\right) &=&\frac{1}{n}\left[ 8Q+2W_{2}-e\right] ,
\label{Fourth Moment Subgraphs} \\
m_{5}\left( A_{\mathcal{G}}\right) &=&\frac{1}{n}\left[ 10\Pi +10\mathcal{C}%
_{dt}-30\Delta \right] .  \label{Fifth Moment Subgraphs}
\end{eqnarray}
\end{lemma}

\medskip

Observe how, as we increase the order of the moments, more complicated
structural features appear in the expressions. In the following example, we
illustrate how to use our expressions to compute the spectral moments of an
online social network from empirical structural data.

\medspace
\begin{figure*}[t]
\centering
\includegraphics[width=1.0\textwidth]{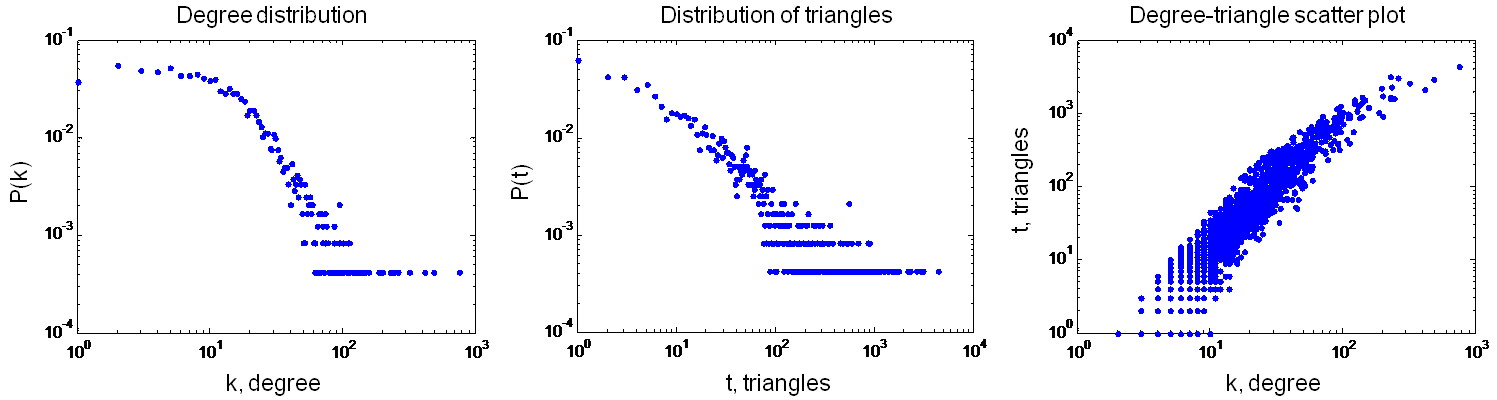}
\caption{In the left and center figures, we plot the distributions of
degrees and triangles of the social network under study (in log-log scale).
In the right figure, we include a scatter plot where each point has
coordinates $(d_{i},t_{i})$, in log-log scale, for all the nodes in the
social graph.}
\label{fig_7}
\end{figure*}

\medskip

\begin{example}
In this example, we study a subgraph of Facebook obtained by exploring the
online social network around a particular node, as follows. From a
particular starting node, we crawl the network (using a breadth-first
search) until we discover the set of all nodes that are within a radius $2$
from the starting node. Using this set of nodes, and their interconnections
(friendships), we construct a social subgraph $\mathcal{F}$ that has $2,404$
nodes and $22,786$ edges. Using this real dataset, we compute the degrees $%
d_{i}$, the number of triangles $t_{i}$, quadrangles $q_{i}$, and pentagons $%
p_{i}$ touching each node $i\in \mathcal{V}\left( \mathcal{F}\right) $. In
Fig. 1, we plot the distributions of degrees and triangles, as well as a
scatter plot of $t_{i}$ versus $d_{i}$ (where each point has coordinates $%
\left( d_{i},t_{i}\right) $, in log-log scale, for all $i\in \mathcal{V}%
\left( \mathcal{F}\right) $). We can aggregate those quantities that are
relevant to compute the spectral moments to obtain the following numerical
values:%
\begin{equation*}
\begin{array}{ll}
e/n=\sum d_{i}/2n=9.478, & \Delta /n=\sum t_{i}/3n=28.15, \\ 
Q/n=\sum q_{i}/4n=825.3, & \Pi /n=\sum p_{i}/5n=31,794, \\ 
W_{2}/n=\sum d_{i}^{2}/n=1,318, & \mathcal{C}_{dt}/n=\sum d_{i}t_{i}/n=8,520.%
\end{array}%
\end{equation*}%
Hence, using Corollary \ref{One corollary} and Lemma \ref{Lemma 4th metrics}%
, we obtain the following values for the spectral moments: $m_{1}\left( A_{%
\mathcal{F}}\right) =0,$ $m_{2}\left( A_{\mathcal{F}}\right) =18.95,$ $%
m_{3}\left( A_{\mathcal{F}}\right) =168.9,$ $m_{4}\left( A_{\mathcal{F}%
}\right) =9,230,$ and $m_{5}\left( A_{\mathcal{F}}\right) =402,310.$
\end{example}

%We close this section with several remarks. First, although one
%could derive closed expressions for moments of order greater than 5, the
%structural pieces of information involved in the moments would be too expensive to determine. For example, in order to compute %the sixth moment, we would need to count the number of hexagons (closed walks of length 6) in the network. This quantity cannot %be locally computed if the node only has access to its 2-hop neighborhood.\footnote{%
%Although other alternatives could be proposed to surpass this issue, such as
%a message-passing approach, we consider only structural metrics that can be 
%\emph{directly} extracted from the 2-hop neighborhoods.} As a general rule, 
%\emph{if nodes have access to their }$r$\emph{-hop neighborhoods, we can
%compute the sequence of spectral moments up to order }$k_{\max}=2r+1$\emph{\
%by aggregating local structural information} \cite{PJ10}. In real-world
%networks, the average size of a \emph{3}-hop neighborhood is usually very
%large. (This is a consequence of the `Six Degrees of Separation' phenomenon,
%since 3-hop neighborhoods have diameter 6.) Hence, the computational cost of
%counting subgraphs in $3$-hop neighborhoods become extremely costly.

\bigskip

In this section, we have derived expressions to compute the first five
spectral moment of $A_{\mathcal{G}}$ from network structural properties. In
the next section, we use semidefinite programming to extract information
regarding eigenvalues of interest from a sequence of spectral moments.

\section{\label{Optimal Bounds}Optimal Spectral Bounds from Spectral Moments}

Here, we introduce an approach to derive an upper bound on the smallest
eigenvalues of $A_{\mathcal{G}}$ from its sequence of spectral moments%
\footnote{%
As a by-product of our analysis, we also derive lower bounds on the spectral
radius of $A_{\mathcal{G}}$, although these bounds are not essential in our
analysis.}. Since we have expressions for the spectral moments in terms of
local structural properties, our bounds relate the eigenvalues of a network
with these properties.\ There is a large literature studying the
relationship between structural and spectral properties of graphs (see \cite%
{CDS80},\cite{DK04}, and references therein, for an extensive list of
spectral results). For many real-world networks, there is a particular set
of structural properties that play a key role in the network's
functionality. For example, it is well-known that social networks contain a
large number of triangles (and other cycles). Hence, it would be useful to
have spectral bounds where these structural features are jointly
represented. In this section, we derive new upper bounds on the smallest
eigenvalue of the adjacency matrix in terms of the structural properties
involved in (\ref{Moments as Averages}), (\ref{Fourth Moment Subgraphs}) and
(\ref{Fifth Moment Subgraphs}). Our results can be easily extended to derive
lower bounds on the spectral radius of the adjacency matrix, although this
bound is not of relevance in our analysis of games with linear best
responses.

Now, we derive bounds on the smallest eigenvalue of the adjacency matrix in
terms of relevant structural properties by adapting the optimization
framework proposed in \cite{Las11}. We first need to introduce a
probabilistic interpretation of a network eigenvalue spectrum and its
spectral moments. For a simple graph $\mathcal{G}$, we define its spectral
density as:%
\begin{equation}
\mu _{\mathcal{G}}\left( x\right) =\frac{1}{n}\sum_{i=1}^{n}\delta \left(
x-\lambda _{i}\right) ,  \label{Spectral Measure}
\end{equation}%
where $\delta \left( \bullet \right) $ is the Dirac delta function and $%
\left\{ \lambda _{i}\right\} _{i=1}^{n}$ is the set of (real) eigenvalues of
the (symmetric) adjacency matrix $A_{\mathcal{G}}$. Consider a random
variable $X$ with probability density $\mu _{\mathcal{G}}$. The moments of $%
X\sim \mu _{\mathcal{G}}$ are equal to the spectral moments of $A_{\mathcal{G%
}}$, i.e.,%
\begin{align*}
\mathbb{E}_{\mu _{\mathcal{G}}}\left( X^{k}\right) & =\int_{\mathbb{R}%
}x^{k}\mu _{\mathcal{G}}\left( x\right) dx \\
& =\frac{1}{n}\sum_{i=1}^{n}\int_{\mathbb{R}}x^{k}\delta \left( x-\lambda
_{i}\right) dx \\
& =\frac{1}{n}\sum_{i=1}^{n}\lambda _{i}^{k}=m_{k}\left( A_{\mathcal{G}%
}\right) ,
\end{align*}%
for all $k\geq 0$.

In \cite{Las11}, Lasserre proposed a technique to compute the smallest
interval $\left[ a,b\right] $ containing the support\footnote{%
Recall that the support of a finite Borel measure $\mu $ on $R$, denoted by $%
supp\left( \mu \right) $, is the smallest closed set $B$ such that $\mu
\left( R\backslash B\right) =0$.} of a positive Borel measure $\mu $ from
its complete sequence of moments $\left( m_{r}\right) _{r\geq 0}$. In our
spectral problem, the positive Borel measure under consideration is the
spectral density $\mu _{\mathcal{G}}\left( x\right) $, defined in (\ref%
{Spectral Measure}). Hence, in the context of our problem, the sequence of
moments $\left( m_{r}\right) _{r\geq 0}$ is equal to $\left( m_{r}\left( A_{%
\mathcal{G}}\right) \right) _{r\geq 0}$, and the smallest interval $\left[
a,b\right] $ containing the support of $\mu _{\mathcal{G}}\left( x\right) $
is equal to $\left[ \lambda _{n},\lambda _{1}\right] $, by the definition in
(\ref{Spectral Measure}).

Lasserre also proposed in \cite{Las11} a numerical scheme to compute tight
bounds on the values of $a$ and $b$ when a \emph{truncated} sequence of
moments $\left( m_{r}\right) _{0\leq r\leq k}$ is known. This numerical
scheme involves a series of semidefinite programs (SDP) in one variable. As
we show below, at step $s$ of this series of SDP's, we are given a sequence
of moments $\left( m_{1},...,m_{2s+1}\right) $ and solve two SDP's whose
solution provides an inner approximation $\left[ \alpha _{s},\beta _{s}%
\right] \subseteq \left[ a,b\right] $. In our case, since we have
expressions for the first five spectral moments, $\left( m_{1}\left( A_{%
\mathcal{G}}\right) ,...,m_{5}\left( A_{\mathcal{G}}\right) \right) $, we
can solve the first two steps of this series of SDP's to find inner
approximations $\left[ \alpha _{s},\beta _{s}\right] \subseteq \left[
\lambda _{n},\lambda _{1}\right] $. In other words, the solution to the
SDP's provide us with the bounds $\alpha _{s}\geq \lambda _{n}$ and $\beta
_{s}\leq \lambda _{1}$.

In order to formulate the series of SDP's proposed in \cite{Las11}, we need
to define the so-called localizing matrix of our problem \cite{LasBOOK}.
Given a sequence of moments, $\mathbf{m}^{\left( 2s+1\right) }=\left(
m_{1},...,m_{2s+1}\right) $, our localizing matrix is a Hankel matrix
defined as:%
\begin{equation}
H_{s}\left( c\right) \triangleq R_{2s+1}-c~R_{2s}\text{,}
\label{Localizing matrix}
\end{equation}%
where $R_{2s}$ and $R_{2s+1}$ are the Hankel matrices of moments defined \ as%
\begin{align}
R_{2s}& =\left[ 
\begin{array}{cccc}
1 & m_{1} & \cdots & m_{s} \\ 
m_{1} & m_{2} & \cdots & m_{s+1} \\ 
\vdots & \vdots & \ddots & \vdots \\ 
m_{s} & m_{s+1} & \cdots & m_{2s}%
\end{array}%
\right] ,  \label{Hankel matrices} \\
R_{2s+1}& =\left[ 
\begin{array}{cccc}
m_{1} & m_{2} & \cdots & m_{s+1} \\ 
m_{2} & m_{3} & \cdots & m_{s+2} \\ 
\vdots & \vdots & \ddots & \vdots \\ 
m_{s+1} & m_{s+2} & \cdots & m_{2s+1}%
\end{array}%
\right] .  \notag
\end{align}%
Hence, for a given sequence of moments, the entries of $H_{s}\left( c\right) 
$ depend affinely on the variable $c$. We can compute $\alpha _{s}$ and $%
\beta _{s}$ as the solution to the following semidefinite programs \cite%
{Las11}:

\begin{proposition}
\label{Lasserre bound}Let $\mathbf{m}^{\left( 2s+1\right) }=\left(
m_{1},...,m_{2s+1}\right) $ be the truncated sequence of moments of a
positive Borel measure $\mu $. Then,%
\begin{align}
a& \leq \alpha _{s}\triangleq \max_{\alpha }\left\{ \alpha :H_{s}\left(
\alpha \right) \succcurlyeq 0\right\} ,  \label{Bound min eigenval} \\
b& \geq \beta _{s}\triangleq \min_{\beta }\left\{ \beta :-H_{s}\left( \beta
\right) \succcurlyeq 0\right\} ,  \label{Bound max eigenval}
\end{align}%
for $\left[ a,b\right] $ being the smallest interval containing $supp(\mu )$.
\end{proposition}

\bigskip

\begin{remark}
Observe that $\alpha _{s}$ and $\beta _{s}$ are the solutions to two SDP's
in one variable, since the constraint $H_{s}\left( \alpha \right)
\succcurlyeq 0$ (resp. $-H_{s}\left( \beta \right) \succcurlyeq 0$)
indicates that the matrix $H_{s}\left( \alpha \right) $ (resp. $-H_{s}\left(
\beta \right) $) is positive semidefinite and this matrix has affine entries
with respect to $\alpha $ (resp. $\beta $). Hence, they can be efficiently
computed using standard optimization software (for example, CVX \cite{CVX}).
As we increase $s$ in Proposition \ref{Lasserre bound}, more moments are
involved in the SDP's, and the resulting bounds become tighter, i.e., $%
\alpha _{s+1}\leq \alpha _{s}$ and $\beta _{s+1}\geq \beta _{s}$.
\end{remark}

In the context of our spectral analysis, the Borel measure in Proposition %
\ref{Lasserre bound} corresponds to the spectral density of a graph $%
\mathcal{G}$, and the smallest interval $\left[ a,b\right] $ corresponds to $%
\left[ \lambda _{n}\left( A_{\mathcal{G}}\right) ,\rho \left( A_{\mathcal{G}%
}\right) \right] $. Thus, Proposition \ref{Lasserre bound}\ provides an
efficient numerical scheme to compute the bounds $\alpha _{s}\geq \lambda
_{n}\left( A_{\mathcal{G}}\right) $ and $\beta _{s}\leq \rho \left( A_{%
\mathcal{G}}\right) $. When we are given a sequence of five spectral
moments, we can solve the SDP's in (\ref{Bound min eigenval}) and (\ref%
{Bound max eigenval}) analytically for $s=2$. In this case, the localizing
matrix is:%
\begin{equation}
H_{2}\left( c\right) =\left[ 
\begin{array}{ccc}
m_{1}-c & m_{2}-cm_{1} & m_{3}-cm_{2} \\ 
m_{2}-cm_{1} & m_{3}-cm_{2} & m_{4}-cm_{3} \\ 
m_{3}-cm_{2} & m_{4}-cm_{3} & m_{5}-cm_{4}%
\end{array}%
\right] .  \label{H2c}
\end{equation}%
As we proved in Section \ref{Spectral Section}, the spectral moments in the
localizing matrix depend on the number of nodes, edges, cycles of length 3
to 5, the sum-of-squares of degrees $W_{2}$, and the degree-triangle
correlation $\mathcal{C}_{dt}$.

Furthermore, for $s=2$, the optimal values $\alpha _{2}$ and $\beta _{2}$
can be analytically computed, as follows. First, note that $%
H_{2}(c)\succcurlyeq 0$ (resp. $-H_{2}(c)\succcurlyeq 0$) if and only if all
the eigenvalues of $H_{2}$ are nonnegative (resp. nonpositive). For a given
sequence of five moments, the characteristic polynomial of $H_{2}\left(
c\right) $ can be written as%
\begin{equation*}
\phi _{2}\left( \lambda \right) \triangleq \det \left( \lambda
I-H_{2}(c)\right) =\lambda ^{3}+p_{1}\left( c\right) \lambda
^{2}+p_{2}\left( c\right) \lambda +p_{3}\left( c\right) ,
\end{equation*}%
where $p_{j}\left( c\right) $ is a polynomial of degree $j$ in the variable $%
c$ (with coefficients depending on the moments). Thus, by Descartes' rule,
all the eigenvalues of $H_{2}\left( c\right) $ are nonpositive if and only
if $p_{j}\left( c\right) \geq 0$, for $j=1,2,$ and $3$. Similarly, all the
eigenvalues are nonnegative if and only if $p_{2}\left( c\right) \geq 0$ and 
$p_{1}\left( c\right) ,p_{3}\left( c\right) \leq 0$. In fact, one can prove
that the optimal value of $\alpha _{2}$ and $\beta _{2}$ in (\ref{Bound max
eigenval}) can be computed as the smallest and the largest roots of $\det
H_{2}\left( c\right) =0$, which yields a third degree polynomial in the
variable $c$ \cite{Las11}. There are closed-form expressions for the roots
of this polynomial (for example, Cardano's formula \cite{AS65}), although
the resulting expressions for the roots are rather complicated.

In this subsection, we have presented a convex optimization framework to
compute optimal bounds on the maximum and the minimum eigenvalues of a graph 
$\mathcal{G}$ from a truncated sequence of its spectral moments. Since we
have expressions for spectral moments in terms of local structural
properties, these bounds relate the eigenvalues of a graph with its
structural properties.

\section{\label{Simulations}Numerical Simulations}

As we illustrated in Section \ref{Strategic Section}, there is a close
connection between the largest and the smallest eigenvalues of a network and
the outcome of a game with linear best response functions. In this section,
we use our bounds on the support of the eigenvalue spectrum to study the
role of structural properties in the existence and the stability of a Nash
equilibrium. For this purpose, we analyze real data from a regional network
of Facebook that spans $63,731$ users (nodes) connected by $817,090$
friendships (edges) \cite{VMCG09}. In order to corroborate our results in
different network topologies, we extract multiple medium-size social
subgraphs from the Facebook graph by running a Breath-First Search (BFS)
around different starting nodes. Each BFS induces a social subgraph spanning
all nodes 2 hops away from a starting node, as well as the edges connecting
them. We use this approach to generate a set $\mathbf{G}=\{G_{i}\}_{i\leq
100}$ of 100 different social subgraphs centered around 100 randomly chosen
nodes.\footnote{%
Although this procedure is common in studying large social network , it
introduces biases that must be considered carefully \cite{SWM05}.}

%In our first simulation, we validate the usage of the spectral radius as a
%measure of the spreading abilities of a social network. We consider $25$
%different social subgraphs, with sizes between $1000$ and $3000$ nodes, and
%study the behavior of viral spreading processes for values of $\delta/\beta$
%equal to $\rho$, $0.85\rho$, and $0.70\rho$. For each value of $\delta/\beta$%
%, we simulate $25$ different  realizations of the viral spreading,
%with a $2\%$ probability of initial infection, and compute the final
%fraction of individuals in the network that have been infected during the
%time of the simulation ($250$ steps). In Fig. 6, we plot the average
%fraction of infected individuals versus $\rho\left( A_{\mathcal{G}}\right) $
%for the three different values of $\delta/\beta$. We observe that for $%
%\delta/\beta =\rho\left( A_{\mathcal{G}}\right) $ the portion of infected
%population lie under $20\%$ for most cases (circles in Fig. 6). For $%
%\delta/\beta =0.70\rho\left( A_{\mathcal{G}}\right) $, the portion of
%infected mostly lies above $80\%$ (squares in Fig. 6). For $%
%\delta/\beta=0.85\rho\left( A_{\mathcal{G}}\right) $, all the points lie in
%the transition band $20\%$-$80\%$.

%\begin{figure}[t]
%\centering
%\includegraphics[width=0.4\textwidth]{Fig9C.jpg}
%\caption{Each point represents the average fraction of eventually infected
%nodes when $ \delta/ \beta$ equals $ \rho$ (circles), $0.85 %
%\rho$ (crosses), and $0.70 \rho$ (squares).}
%\label{fig_9}
%\end{figure}

\begin{figure}[t]
\centering
\includegraphics[width=0.43\textwidth]{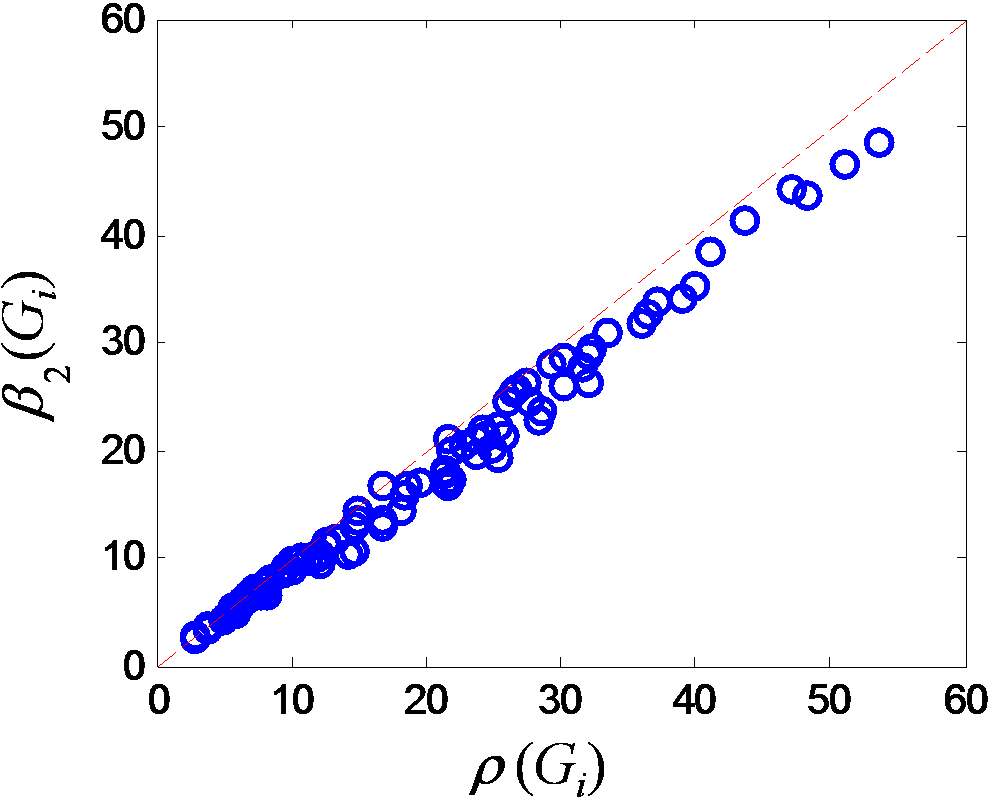}
\caption{Scatter plot of the spectral radius, $\protect\rho \left(
G_{i}\right) $, versus the lower bound $\protect\beta _{2}\left(
G_{i}\right) $ , where each point is associated with one of the $100$ social
subgraphs considered in our experiments.}
\end{figure}

From Corollary \ref{One corollary} and Lemma \ref{Lemma 4th metrics} we can
compute the first five spectral moments of a graph $G_{i}$ from the
following structural properties: number of nodes ($n_{i}$), edges ($e_{i}$),
triangles ($\Delta _{i}$), quadrangles ($Q_{i}$), pentagons ($\Pi _{i}$), as
well as the sum-of-squares of the degrees ($W_{i}$), and the degree-triangle
correlation ($\mathcal{C}_{i}$). For convenience, we define $S\left(
G_{i}\right) \triangleq \left\{ n_{i},e_{i},\Delta _{i},Q_{i},\Pi _{i},W_{i},%
\mathcal{C}_{i}\right\} $ as a set of relevant structural properties of $%
G_{i}$. In our numerical experiment, we first measure the set of relevant
properties $S_{i}$ for each social subgraph $G_{i}\in \mathbf{G}$, and then
compute the first five spectral moments of its adjacency matrix. From these
moments, we then compute the bounds $\alpha _{2}\left( G_{i}\right) $ and $%
\beta _{2}\left( G_{i}\right) $ using Proposition \ref{Lasserre bound}. As
we mentioned before, these bounds can be computed as the maximum and minimum
roots of a third order polynomial, for which closed form expressions are
known.

We illustrate the quality of our bounds in the following figures. Fig. 2 is
a scatter plot where each circle has coordinates $\left( \rho \left(
G_{i}\right) ,\beta _{2}\left( G_{i}\right) \right) $, for all $G_{i}\in 
\mathbf{G}$. Observe how \emph{the spectral radii }$\rho \left( G_{i}\right) 
$ \emph{of these social subgraphs are remarkably close to the theoretical
lower bound} $\beta _{2}\left( G_{i}\right) $. Therefore, we can use $\beta
_{2}\left( G_{i}\right) $ as an estimate of $\rho \left( G_{i}\right) $ for
social subgraphs. In Fig. 3 we include a scatter plot where each circle has
coordinates $\left( -\lambda _{n}\left( G_{i}\right) ,-\alpha _{2}\left(
G_{i}\right) \right) $, for all $G_{i}\in \mathbf{G}$. Although $\alpha
_{2}\left( G_{i}\right) $ is a looser bound than $\beta _{2}\left(
G_{i}\right) $, we observe how there is a strong correlation between the
value of $\lambda _{n}\left( G_{i}\right) $ and $\alpha _{2}\left(
G_{i}\right) $.

In these numerical experiments, we have first showed that $\alpha _{2}\left(
G_{i}\right) $ and $\beta _{2}\left( G_{i}\right) $ bound the smallest and
the largest eigenvalues of the adjacency matrix, and that these bounds are
tight, specially $\beta _{2}\left( G_{i}\right) $. Since these bounds can be
written as explicit functions of the structural properties in $S\left(
G_{i}\right) $, we can estimate the impact of structural perturbations on
the spectral radius and the smallest eigenvalue by studying $\left. \partial
\beta _{2}\right/ \partial p_{i}$ and $\left. \partial \alpha _{2}\right/
\partial p_{i}$ for $p_{i}\in S\left( G_{i}\right) $. (Details of this
perturbation analysis are left for future work due to space limitations.)

\bigskip

\begin{figure}[t]
\centering
\includegraphics[width=0.43\textwidth]{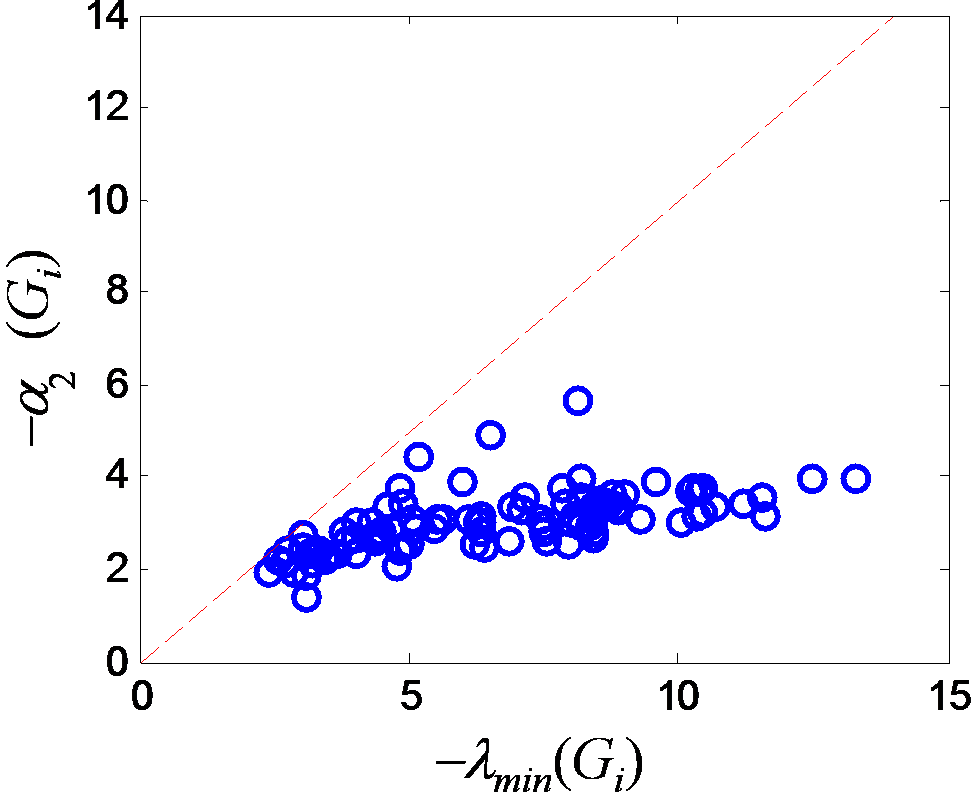}
\caption{Scatter plot of $-\protect\lambda _{\min }\left( G_{i}\right) $
versus the bound $-\protect\alpha _{2}\left( G_{i}\right) $ , where each
point is associated with one of the $100$ social subgraphs considered in our
experiments.}
\end{figure}

\section{Conclusions}

In this paper, we have studied games with linear best response functions in
a networked context. We have focused on analyzing the role of the network
structure on the game outcome. In particular, the existence and the
stability of a unique Nash equilibrium in this class of games are closely
related with the smallest eigenvalue of the adjacency matrix of the network.
We take this spectral result as the foundation to our work, and use
algebraic graph theory and convex optimization to study how local structural
properties of the network affect this eigenvalue. In particular, we have
derived expressions for the first five spectral moments of the adjacency
matrix in terms of local structural properties. These structural properties
are: the number of nodes and edges, the number of cycles of length up to 5,
the sum-of-squares of the degrees, and the degree-triangle correlation. From
this sequence of five spectral moments, we propose a novel methodology to
compute optimal bounds on the smallest and the largest eigenvalues of the
adjacency matrix by solving two semidefinite programs. In our case, we are
able to find analytical solutions to these optimal bounds by computing the
roots of a cubic polynomial, for which closed-form expressions are
available. Finally, we have verified the quality of our bounds by running
numerical simulations in a set of 100 online social subgraphs. For future
work, we shall use the results herein presented to study the effect of
structural perturbations in the relevant eigenvalues of the adjacency
matrix, and in properties of the Nash equilibrium.


\begin{thebibliography}{99}
\bibitem{FT91} D. Fudenberg and J. Tirole, \emph{Game Theory}, MIT Press,
1991.

\bibitem{AP07} G.-M. Angeletos and A. Pavan, \textquotedblleft Efficient Use
of Information and Social Value of Information,\textquotedblright\ \emph{%
Econometrica}, vol. 75, pp. 1103-1142, 2007.

\bibitem{BCZ06} C. Ballester, A. Calv\'{o}-Armengol, and Y. Zenou,
\textquotedblleft Who%
%TCIMACRO{\U{b4}}%
%BeginExpansion
\'{}%
%EndExpansion
s Who in Networks. Wanted: The Key Player,\textquotedblright\ \emph{%
Econometrica}, vol 74, pp. 1403-1417, 2006.

\bibitem{BK07} Y. Bramoull\'{e} and R. Kranton, \textquotedblleft Public
Goods in Networks,\textquotedblright\ \emph{Journal of Economic Theory},
vol. 135, pp. 478-494, 2007.

\bibitem{BKA10} Y. Bramoull\'{e}, R. Kranton, and M. D'Amours,
\textquotedblleft Strategic Interaction and Networks,\textquotedblright\ 
\emph{CIRPEE Working Paper} 10-18, 2010.

\bibitem{BC07} C. Ballester and A. Calv\'{o}-Armengol, \textquotedblleft
Moderate Interactions in Games with Induced
Complementaries,\textquotedblright\ mimeo, Universidad Aut\'{o}noma de
Barcelona, 2007.

\bibitem{Big93} N. Biggs, \emph{Algebraic Graph Theory}, Cambridge
University Press, 2$^{nd}$ Edition, 1993.

\bibitem{PJ10} V.M. Preciado and A. Jadbabaie, \textquotedblleft From Local
Measurements to Network Spectral Properties: Beyond Degree
Distributions,\textquotedblright\ \emph{Proc. IEEE Conference on Decision
and Control}, 2010.

\bibitem{S01} S.H. Strogatz, \textquotedblleft Exploring Complex
Networks,\textquotedblright\ \emph{Nature}, vol. 410, pp. 268-276, 2001.

\bibitem{BLMCH06} S. Boccaletti S., V. Latora, Y. Moreno, M. Chavez, and
D.-H. Hwang, \textquotedblleft Complex Networks: Structure and
Dynamics,\textquotedblright\ \emph{Physics Reports}, vol. 424, no. 4-5, pp.
175-308, 2006.

\bibitem{Pre08} V.M. Preciado, \emph{Spectral Analysis for Stochastic Models
of Large-Scale Complex Dynamical Networks}, Ph.D. dissertation, Dept. Elect.
Eng. Comput. Sci., MIT, Cambridge, MA, 2008.

\bibitem{CDS80} D. Cvetkovi\'{c}, M. Doob and H. Sachs, \emph{Spectra of
Graphs}, Wiley-VCH, $3^{rd}$ Edition, 1998.

\bibitem{DK04} K.C. Das and P. Kumar, \textquotedblleft Some New Bounds on
the Spectral Radius of Graphs,\textquotedblright\ \emph{Discrete Mathematics}%
, vol. 281, pp. 149-161, 2004.

\bibitem{Las11} J.B. Lasserre, \textquotedblleft Bounding the Support of a
Measure from its Marginal Moments,\textquotedblright\ \emph{Proc. AMS}, in
press.

\bibitem{LasBOOK} J.B. Lasserre, \emph{Moments, Positive Polynomials and
Their Applications}, Imperial College Press, London, 2009.

\bibitem{CVX} $<$ http://cvxr.com/cvx/$>$

\bibitem{AS65} M. Abramowitz and I.A. Stegun, \emph{Handbook of Mathematical
Functions with Formulas, Graphs, and Mathematical Tables}, Dover, 1965.

\bibitem{VMCG09} B. Viswanath, A. Mislove, M. Cha, and K.P. Gummadi,
\textquotedblleft On the Evolution of User Interaction in
Facebook,\textquotedblright\ \emph{Proc. ACM SIGCOMM Workshop on Social
Networks}, 2009.

\bibitem{SWM05} M. Stumpf, C. Wiuf, and R. May, \textquotedblleft Subnets of
Scale-Free Networks are not Scale-Free: Sampling Properties of
Networks,\textquotedblright\ \emph{Proceedings of the National Academy of
Sciences}, vol. 102, pp. 4221-4224, 2005.
\end{thebibliography}
\end{document}